\documentclass[aip, twocolumn, reprint, cha]{revtex4-1}
\usepackage[utf8]{inputenc}
\usepackage[T1]{fontenc}
\usepackage[english]{babel}
\usepackage{dcolumn} 
\usepackage{graphicx}
\usepackage{latexsym}
\usepackage{pstricks, pst-node, pst-text, pst-3d}
\usepackage{amssymb, amsmath, amscd}
\usepackage{mathrsfs, mathtools}
\usepackage{verbatim}
\usepackage{hyperref}
\usepackage{placeins}
\usepackage{multirow}
\usepackage{pifont}
\usepackage{lipsum}
\usepackage{array}
\usepackage{color}
\usepackage{bm}
\hypersetup{
	colorlinks = true, %colors links instead of ugly boxes
	urlcolor   = magenta, %color for external hyperlinks
	linkcolor  = red, %color of internal links
	citecolor  = blue %color of citations
}
\definecolor{green}{rgb}{0,0.5,0}
\definecolor{teal}{rgb}{0,0.7,0.8}
\definecolor{vermelho}{rgb}{1,0.02,0.05}
\definecolor{atencao}{rgb}{1,0.5,0.05}

\begin{document}
	\title{Multistability and chaos in SEIRS epidemic model with a periodic time-dependent transmission rate}
%
%==================================================================================================
%
	\author{Eduardo L. Brugnago}\email{elb@if.usp.br}
	\affiliation{Physics Institute, University of S\~ao Paulo, 05508-090, S\~ao Paulo, SP, Brazil.}	
	\author{Enrique C. Gabrick}\email{ecgabrick@gmail.com} 
	\affiliation{Graduate Program in Science, State University of Ponta Grossa, 84030-900, Ponta Grossa, PR, Brazil.}
	\author{Kelly C. Iarosz}
	\affiliation{Physics Institute, University of S\~ao Paulo, 05508-090, S\~ao Paulo, SP, Brazil.}
	\affiliation{University Center UNIFATEB, 84266-010, Tel\^emaco Borba, PR, Brazil.}
	\affiliation{Santa Helena Institute, 84266-010, Tel\^emaco Borba, PR, Brazil.}
	\author{Jos\'e D. Szezech Jr.}
	\affiliation{Graduate Program in Science, State University of Ponta Grossa, 84030-900, Ponta Grossa, PR, Brazil.}
	\affiliation{Mathematics and Statistics Department, State University of Ponta Grossa, 84030-900, Ponta Grossa, PR, Brazil.}
	\author{Ricardo L. Viana}
	\affiliation{Physics Institute, University of S\~ao Paulo, 05508-090, S\~ao Paulo, SP, Brazil.}	
	\affiliation{Department of Physics, Federal University of Paran\'a, 81531-980, Curitiba, PR, Brazil}
	\author{Antonio M. Batista}
	\affiliation{Physics Institute, University of S\~ao Paulo, 05508-090, S\~ao Paulo, SP, Brazil.}
	\affiliation{Graduate Program in Science, State University of Ponta Grossa, 84030-900, Ponta Grossa, PR, Brazil.}
	\affiliation{Mathematics and Statistics Department, State University of Ponta Grossa, 84030-900, Ponta Grossa, PR, Brazil.}
	\author{Iber\^e L. Caldas}
	\affiliation{Physics Institute, University of S\~ao Paulo, 05508-090, S\~ao Paulo, SP, Brazil.}	
%	
%==================================================================================================
%	
	\begin{abstract}
		In this work, 
		we study the dynamics of a SEIRS epidemic model with a periodic time-dependent transmission rate.  
		Emphasizing the influence of the seasonality frequency on the system dynamics, 
		we analyze the largest Lyapunov exponent 
		%of trajectories obtained 
		along parameter planes finding large chaotic regions. 
		Furthermore, 
		in some ranges there are shrimp-like periodic strutures. 
		We highlight the system multistability, 
		identifying the coexistence of periodic orbits for the same parameter values, 
		with the infections maximum distinguishing by up one order of magnitude, 
		depending only on the initial conditions. 
		In this case, 
		the basins of attraction has self-similarity. 
		Parametric configurations, 
		for which both periodic and 
		non-periodic orbits occur, 
		cover $13.20\%$ of the evaluated range. 
		We also identified the coexistence of periodic and 
		chaotic attractors with different maxima of infectious cases, 
		where the periodic scenario peak reaching approximately $50\%$ higher than the chaotic one. 
	\end{abstract}
%
%==================================================================================================
%	
	\maketitle
%
%==================================================================================================
%	
	\textbf{Seasonality is a factor that influences many infections spread. 
		Namely, 
		a time-dependent transmission rate leads to a non-autonomous differential equations system, 
		enriching the dynamics and 
		bringing features not observed in the autonomous models. 
		Causes beyond weather seasons, 
		as control measures and 
		various environmental factors, 
		may be related to temporal variations in the transmission rate of infections. 
		It is known that the seasonality degree is relevant to change the epidemic system dynamics, 
		as well as the average transmissivity. 
		In addition, 
		some diseases have distinct seasonality, 
		as annual (e.g., rubella and measles), 
		biannual (e.g., chikenpox) and 
		irregular (e.g., mumps) peaks. 
		In a general picture, 
		the seasonality frequency does not need to be linked to annual cycles. 
		In this way, 
		we investigate how the seasonality parameters affects the SEIRS model dynamics, 
		emphasizing the role of seasonality frequency. 
	}
%
%==================================================================================================
%
\section{INTRODUCTION} \label{sec_introducao}
{Mathematical models are a fundamental tool to understand the epidemic dynamics~\cite{Keeling2008},  
	{where approximations are made in order to replicate the focus behavior and 
		provide a better understanding.} 
%	In deterministic formulations, 
%	these models depend on the constant parameters and 
%	their solutions are transient going to a steady state equilibrium~\cite{Bjornstad2018}. 
%	However, 
	Several diseases present periodic outbreaks~\cite{Altizer2006}, 
	that are related to non-constant transmission rates~\cite{Grassly2006}. 
	In this way, 
	deterministic systems with constant transmission rate is not realistic~\cite{Greenhalgh2003}. 
	In addition to the better description of real data, 
	the inclusion of the non-constant term leads to chaotic solutions~\cite{Grenfell1995}. 
	{Non-autonomous epidemic models, 
		with periodic transmission rates, 
		make it possible to model the behavior of various seasonal diseases, 
	and present rich dynamics~\cite{aron1984,schwartz1983,kuznetsov1994}.}
	The chaotic solutions have connection with reported data. 
	For example,} 
time series of many epidemic diseases as measles~\cite{Olsen1990}, 
dengue~\cite{Aguiar2011}, 
mumps~\cite{London1973} and 
others~\cite{Keeling2001}, 
can be chaotic. 
%The implications of chaotic regimes raise relevant questions in epidemiology~\cite{Rand1991}. 
%In some of these, % diseases, 
%chaos appears associated with seasonality~\cite{Aguiar2009}. 
%being such feature simulated in epidemiological models through the inclusion of a non-linear term 
Such behavior is associated with the seasonality present in recurrent infections~\cite{Aguiar2009,Tanaka2013}. 
%Also, 
%seasonality behavior is present in recurrent diseases~\cite{Aguiar2009,Tanaka2013}. 
In order to simulate the chaotic dynamic, 
a non-linear term is included in the equations~\cite{Bjornstad2018, Altizer2006}, 
which can be given, 
for example, 
by a square wave or sine function~\cite{Tanaka2013}. 
The implications of chaotic regimes raise relevant questions in epidemiology~\cite{Rand1991}, 
where a consequence of the chaotic dynamics is a reduction of forecast horizons for new outbreaks~\cite{Keeling2008}.  
%being more impactful with the time series length increasing~\cite{Scarpino2019}. 
From a mathematical modeling perspective, 
many of the classic epidemiological models have a compartmental structure, 
distributing the host population into classes according to considered stages of the disease spread evolution. 
Essentially, 
there is a compartment for the population susceptible to infection, 
identified by the variable $S$, 
in addition to another one for infectious individuals, 
identified by $I$. 
There may be several other compartments, 
adapting the model to the studied disease characteristics~\cite{AndersonMay,Gabrick2022_PhysA}. 
It is assumed these groups of individuals homogeneously distributed in the population, 
as well as concentrations with different contagion probabilities do not occur. 
Homogeneity of mean disease characteristics, 
such as duration of infectious and 
latency, 
is also considered. 
The mathematical description of a compartmental epidemiological model encompasses both: 
host population subdivision and 
transition rules between the disease stages. 
Secondary infections usually are described by means of an interaction term between populations in $S$ and 
$I$ compartments. 

Since the seminal work of Kermack and 
McKendrick~\cite{Kermack1927}, 
these models have been employed to study many diseases 
spread~\cite{Earn2000, Aguiar2008, Ho2019, Amaku2021, Manchein2020, Brugnago2020, Amelia2022, Dalal2008, Dushoff2004, Galvis2022, mugnaine2022_CLF}. 
Formulations closer to the original proposals do not produce chaotic dynamics~\cite{Cooper2020, Silvio2021, batista2021_RBEF}, 
however inclusion of multistrain~\cite{Bianco2009} or seasonal~\cite{Yi2009} terms are able to reproduce chaos in a wide parameters range. 
Considering a SIR seasonal forced model, 
Stollenwerk et al.~\cite{Stollenwerk2022} investigated the dynamics of respiratory diseases, 
like influenza. 
In this case, 
the seasonality is related to the winter months. 
In their results, 
they found a route to chaotic dynamics via period doubling bifurcations as a function of seasonality degree, 
similarly to the bifurcation cascade found by Yi et al.~\cite{Yi2009} in a SEIR forced model. 
However, 
in the second case, 
additionally to the chaotic dynamics, 
the authors obtained hyperchaotic solutions for some parameter ranges. 
Completely, 
they investigated the dynamical behaviour by Poincar\'e sections and 
parameter planes assessing Lyapunov exponents. 
Another way to enrich the dynamics is by the time dependent modullation of the transmission rate~\cite{Bilal2016}, 
where the system exhibit multistability, 
by the coexistence of chaotic and periodic attractors for some seasonality degrees. 

Bifurcation cascades as a function of various parameters in compartmental models are found in other works. 
Considering a SIR forced model with multistrain, 
Kamo and 
Sasaki~\cite{Kamo2002} showed that the cross-immunity exerts significant influence to the dynamical behaviour. 
For two strain, 
multiples attractors coexists, 
from which the population can switch by the introduction of small random noise in seasonally transmission. 
However, the complex dynamics also is present in models with one strain. 
In a SEIRS seasonal forced framework, 
Gabrick et al.~\cite{Gabrick2023} showed multistable dynamics between chaotic and 
periodic attractors. 
To evidence this, 
they generated hysteresis-type bifurcation diagrams as a function of the recovery and 
average contact rates, 
seasonality degree, 
inverse of immunity and latent periods. 
Numerical simulations showed coexistence of chaotic and 
periodic attractors depending on the parameter range. 
Furthermore, 
investigating the dynamical behaviour as function of the inverse of latent period, 
it is possible to associate critical transitions with tipping points~\cite{Medeiros2017}. 
Once crossed this threshold, 
the spread diseases becomes chaotic. 

A common characteristic of the works mentioned above is a period doubling route to chaos given as function of the seasonality degree. 
However, 
the authors did not take into account the effects of varied seasonality frequency. 
Usually, 
seasonality is attributed to environmental factors and, 
as expected, 
it is very common to be related to the weather seasons throughout the year~\cite{Altizer2006}. 
These seasonal forcings lead to oscillations in the infection transmission rate, 
being conditioned by changes in the contact rate between infectious and 
susceptible individuals, 
the circulation of infectious agents and 
their infectiousness. 
In this study, 
we consider different frequencies for a seasonality function, 
focusing on the influence of this parameter to the system dynamics. 
This easing of the oscillation frequency beyond the usual seasonality, 
together with its amplitude, 
allows to model diverse disturbances in the infection spread. 
%Specifically,
%we investigate the SEIRS model with a time-dependent transmission rate as described. 

Firstly, 
we obtain the disease-free (DFE) solution, 
which is defined by the infection eradication in the host population. 
The stability of this solution is associated with the {\it basic reproduction ratio}~\cite{AndersonMay} ($\mathcal{R}_0$). 
Due to the non-autonomous nature of the differential equations system that describes the model, 
$\mathcal{R}_0$ is also time-dependent and 
oscillate between two extremes, 
bounded according to seasonality degree. 
In the autonomous case, 
for $\mathcal{R}_0<1$ the infection is extinguished and, 
otherwise if $\mathcal{R}_0>1$, 
then the infection grows spreading in the host population. 
However, 
for non-autonomous epidemiological systems, 
this criterion is not directly applicable, 
requiring additional and 
more sophisticated evaluations~\cite{Seir_Sazonal} to determine the DFE solution stability. 
To analyze the system dynamics~\cite{TamasTel}, 
we compute the Lyapunov spectrum along parameter planes. 
%composed by the seasonality frequency and 
%other parameter of the model. 
Our results shows a wide range of chaotic behavior in all four evaluated planes. 
In some regions there are shrimp-like periodic structures~\cite{Santos2016} immersed in the chaotic bands. 
In addition, 
we find multistability.  
%Given this characteristic, 
The orbit resulting from the system evolution depends on the initial conditions~\cite{Feudel1996a, Feudel1997, Feudel2008}. 
%Because of that, 
Even in periodic scenarios, 
the predictability on the final state is hard to be determined~\cite{Grebogi1983}. 
Also, 
for the same parametric configurations, 
we highlight the coexistence of periodic and 
chaotic attractors in scenarios where the maximum number of infectious is higher in the periodic ones. 
%In this sense, 
%to understand these kind of system is very important in epidemiology~\cite{Wei2021}. 

This article is organized as follows:
In Sec.~\ref{sec_modelo}, 
we explain the SEIRS model and 
the system formulation by its variables and 
parameters. 
After, 
we develop a normalized version of it and 
include the seasonality term, 
which is used throughout the study.
We also present the analytical DFE solution and 
the time-dependent basic reproduction ratio as a function of the system parameters. 
In Sec.~\ref{sec_resultados}, 
we present numerical results, 
starting with an analysis of Lyapunov exponents along parameter planes, 
following to attraction basins evidencing the multistability. 
Complementarily, 
we compute the proportion of initial conditions that evolve to periodic dynamics along the parameter plane formed by the average transmission rate and 
the seasonality frequency. 
Finally, 
Sec.~\ref{sec_conclusoes} is devoted to a brief summary of the main results and 
our conclusions.  
%
%==================================================================================================
%
%\newpage
\section{MODEL} \label{sec_modelo}
\begin{figure}[!h]
	\centering
	\includegraphics[width=1.\columnwidth]{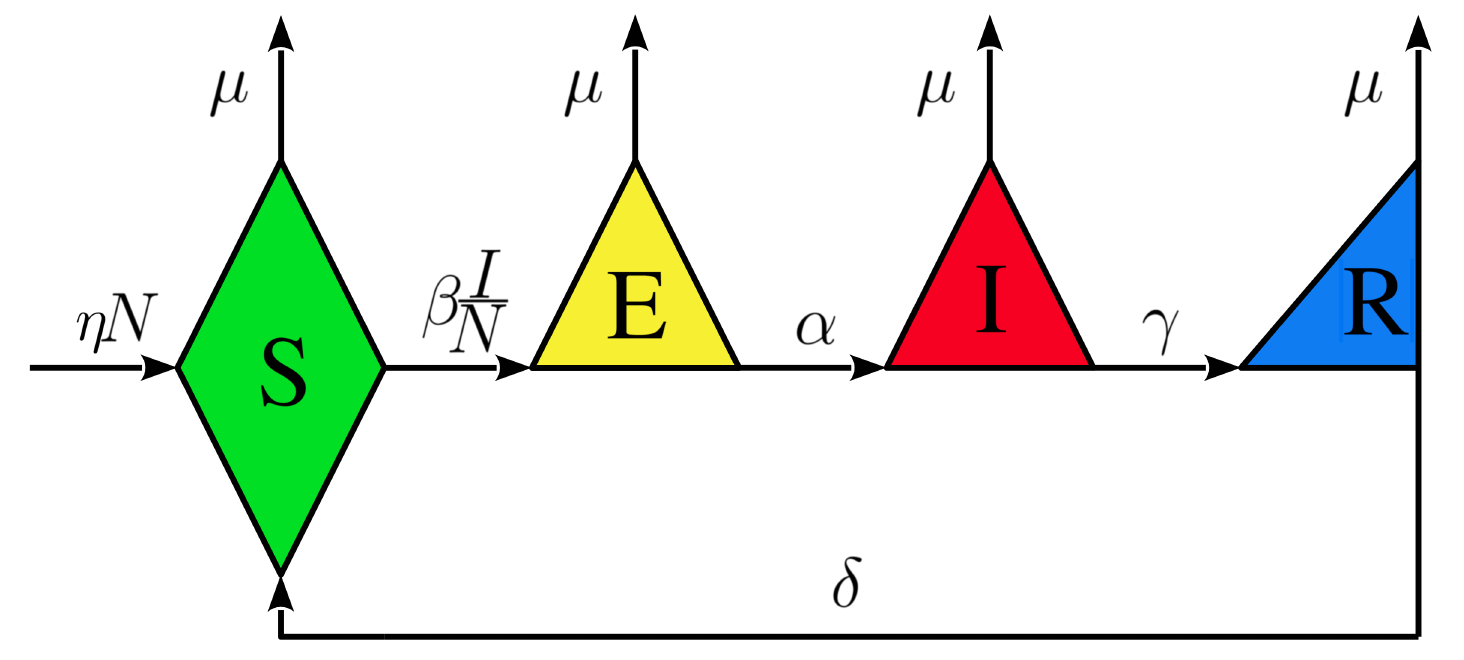}
	\caption{Illustrative scheme of the SEIRS model. 
		The host population is divided into compartments according to the stages of infection. 
		System~(\ref{eq_seirOriginal}) dynamic variables refer to states: 
		susceptible ($S$ in green diamond); 
		exposed ($E$ in yellow triangle); 
		infectious ($I$ in red triangle); 
		and recovered ($R$ in the blue triangle). 
		Arrows connecting compartments indicate the direction of transition between them,  
		with the respective rates. 
		Natural mortality ($\mu$) just decreases the population allocated to each compartment.} 
	\label{fig_modelo}
\end{figure}

Typically presented as a system of four coupled first-order ordinary differential equations, 
SEIRS model describes the disease spread in a host population 
%divided into four compartments, 
%subdivided into four categories (compartments), 
subdivided into four compartments, 
which are identified as the dynamic variables $S$, 
$E$, 
$I$ and 
$R$, 
each corresponding to a portion of the population at different infection stages.  
Given in the form~\cite{Rock2014,Bjornstad2020} 
\begin{align}
	\begin{split}
		\frac{dS}{dt} &= \eta N + \delta R -\left( \beta \frac{I}{N} + \mu \right) S, \\
		\frac{dE}{dt} &= \beta \frac{SI}{N} - (\alpha + \mu) E, \\
		\frac{dI}{dt} &= \alpha E - (\gamma + \mu) I,\\
		\frac{dR}{dt} &= \gamma I - (\delta + \mu) R,
	\end{split}
	\label{eq_seirOriginal}
\end{align}
\noindent this system models the population transitions between its compartments. 
%according to the stages considered in epidemic dynamics. 
Note that the contagion occurs only through interaction between infectious (compartment $I$) and 
susceptible (compartment $S$) populations. 
In order to obtain epidemic scenarios, 
it is necessary to consider an initial condition with already exposed or infected individuals. 
A latency interval is considered, 
for which a portion of the population exposed (compartment $E$) to the pathogen is not yet capable to spread the infection. 
Once an infectious period has elapsed, 
individuals in $I$ acquire temporary immunity %and 
%cease to be infectious, 
passing to the recovered compartment ($R$), 
where they remain until become susceptible to infection again.  
%returning to $S$. 
The host population size is the sum $N = S + E + I + R$. 
An illustrative scheme of this transition dynamics between compartments is shown in Fig.~\ref{fig_modelo}. 

The six parameters of the model correspond to~\cite{Rock2014,Bjornstad2020,Seir_Sazonal}: 
host population birth rate ($\eta$); 
natural death rate ($\mu$); 
%also called general mortality; 
transmission rate ($\beta$); 
mean latent time after the contagion ($1/\alpha$); 
mean infectious period ($1/\gamma$), 
where $\gamma$ is known as recovery rate; 
and mean duration of immunity consequent to infection ($1/\delta$). 
%\atencao{In this study, 
%we assume that the death rate due to infection is negligible compared to overall mortality.} 
All parameters are non-negative real numbers. 
\subsection{System normalization and inclusion of seasonality} \label{subsec_normalizacaoSazonalidade}
We seek a mathematical formulation of the model that allows us to study its dynamics independently of the population size. 
Then, 
without loss of generality, 
we perform the following transformation of variables~\cite{Greenhalgh1997,SEIR_Simplifica}: 
\begin{equation}
	S \mapsto Ns;~~E \mapsto Ne;~~I \mapsto Ni;~~R \mapsto Nr.
	\label{eq_normalizacao}
\end{equation}
where lowercase ones describe the normalized quantities and 
the total population $N>0$. 
Adding the four equations of the system~(\ref{eq_seirOriginal}), 
by means of algebraic manipulation, 
we obtain an expression for the exponential growth of the host population, 
being  
\begin{equation}
	\frac{dN}{dt} = (\eta - \mu)N. 
	\label{eq_populacaoExponencial}
\end{equation}
From this fact and 
the relations proposed in~(\ref{eq_normalizacao}), 
{considering a generic variable $X=Nx$, 
	we obtain}
{ 
\begin{align}
	\frac{dX}{dt} = \frac{d(Nx)}{dt} = N\frac{dx}{dt} + x\frac{dN}{dt}. 
\end{align}
Thus,} 
we get the following transformation for the time derivatives: 
\begin{align}
	\frac{dX}{dt} &= N\left\{\frac{dx}{dt} + (\eta - \mu)x\right\}, \label{eq_transformacao1}\\
	\frac{dx}{dt} &= \frac{1}{N}\left\{\frac{dX}{dt}  - (\eta - \mu)X\right\}, 
\end{align}
where the uppercase $X$ is a non-normalized system variable and 
the lowercase $x$ is its normalized counterpart. 
Operating these transformations and, 
additionally, 
taking into account the constraint $s + e + i + r = 1$, 
we reduce the model to a system of three equations, 
given by 
\begin{align}
	\begin{split}
		\frac{ds}{dt} &= (\delta + \eta)(1 - s) -\beta si  - \delta (e  + i) , \\
		\frac{de}{dt} &= \beta si - (\alpha + \eta) e, \\
		\frac{di}{dt} &= \alpha e - (\gamma + \eta) i, 
	\end{split}
	\label{eq_seirNormalizado}
\end{align}
whose form is identical to that arising from the constant population approximation~\cite{Gabrick2023} ($\eta=\mu$). 

Dynamic variables normalized according to~(\ref{eq_normalizacao}) 
%represent proportions of the amounts in each compartment and 
%the host population, 
represent fractions of the host population in each compartment, 
such that $s,e,i,r\in[0,1]$, 
respecting the constraint between them. 
Thus, 
the system~(\ref{eq_seirNormalizado}) allows simulating epidemic dynamics even in non-constant population scenarios~\cite{SEIR_Simplifica}. 
%Aiming for asymptotic analyzes,
%$\eta - \mu \geq 0$ must be checked. 

In order to model a seasonal behavior of the transmission rate, 
we replace the constant parameter $\beta$ by the periodic function~\cite{Olsen1990} 
\begin{equation}
	\beta(t) = \beta_0 \left[ 1 + \beta_1 \cos (\omega t)  \right], 
	\label{eq_betaPeriodica}
\end{equation}
such that $\beta(t)$ oscillates sinusoidally around the average transmission $\beta_0$, 
with a peak-to-peak variation equal to $2\beta_0 \beta_1$.
Throughout the text, 
we refer to $\beta_1$ as seasonality degree. 
To preserve the meaning of the epidemiological model, 
$\beta(t) \geq 0$ is required, 
hence $\beta_0 \geq 0$ and $\beta_1 \in [0,1]$. 
In this study, 
we did not investigate the case $\beta_0 = 0$, 
since there is no spread of infection. 
Still, 
interested in the seasonality effects, 
for numerical simulations we consider $0 < \beta_1 \leq 1$ and 
$\omega > 0$. 
We extend the idea of seasonality to periodic oscilations not necessarily corresponding to weather seasons, 
not even equivalent to integer multiples or submultiples of one year. 
Here, 
we use this term in a broader sense, 
referring to periodic oscillating transmission rates with any frequency. 
\subsection{Disease-free equilibrium} \label{sec_pontoFixo}
Disease-free equilibrium is so named to signify the disease disappearance in the host population. 
Assuming the fixed point condition 
\begin{equation}
	\frac{ds}{dt}\bigg|_{P_{\rm DFE}} = 
	\frac{de}{dt}\bigg|_{P_{\rm DFE}} = 
	\frac{di}{dt}\bigg|_{P_{\rm DFE}} = 0, 
\end{equation}
at the point $P_{\rm DFE}(s_*,e_*,i_*)$ to the system~(\ref{eq_seirNormalizado}), 
we can obtain the DFE solution~\cite{AndersonMay} directly %from the resulting equations
solving the equations 
\begin{align}
	0 &= (\delta + \eta)(1 - s_*) -\beta s_*i_*  - \delta (e_*  + i_*), \label{eq_sZero}\\
	0 &= \beta s_*i_* - (\alpha + \eta) e_*, \label{eq_eZero} \\
	0 &= \alpha e_* - (\gamma + \eta) i_*. \label{eq_iZero}	
\end{align}

We have that $s_*$, 
$e_*$ and 
$i_*$ are coordinates of the fixed point, 
therefore constant, 
and $\beta = \beta(t) = 0$ if and only if $\beta_1 = 1$ and 
$t = (2 k + 1) \pi/ \omega$, 
with $k \in \mathbb{Z}$ and $\beta_0,\omega > 0$. 
Combining these facts with both Eqs.~(\ref{eq_sZero}) and 
(\ref{eq_eZero}) implies $s_*i_* = 0$ and 
results $e_* = i_* = 0$ in Eqs.~(\ref{eq_eZero}) and 
(\ref{eq_iZero}), 
indicating the infection extinction. 
As for $s_*$, 
there are two distinct cases: 
the first for $\delta + \eta > 0$, 
which gives $s_*=1$ in Eq.~(\ref{eq_sZero}); 
the second occurs if $\delta = \eta = 0$, 
in this case $s_* \in [0,1]$ depends on the initial conditions. 
The second case is more restrictive: 
if $\eta = 0$ the model does not consider births and 
the population in non-normalized system~(\ref{eq_seirOriginal}) decays exponentially with the rate $\mu$, 
according to Eq.~(\ref{eq_populacaoExponencial}); 
while $\delta = 0$ represents that infected individuals acquire permanent immunity after the infectious period, 
reducing the system to a SEIR model. 
{Note that for $\beta$ as a function of time, 
	DFE is the only fixed-point solution.} 

The stability of this solution is related to the basic reproduction ratio $\mathcal{R}_0$~\cite{AndersonMay}. 
For autonomous systems, 
this is a simple linear stability analysis based on the eigenvalues of the system's Jacobian matrix calculated in the DFE point. 
However, 
in this non-autonomous case $\mathcal{R}_0$ is time dependent and 
the eradication of infection may depend of the maximum value  $\max_t\left\{ \mathcal{R}_0(t) \right\}$ or, 
specifically, 
related with the basic reproduction ratio obtained for the long-term average system~\cite{Seir_Sazonal}. 
\subsection{Basic Reproduction Ratio} \label{sec_r0}
The contagion rate is associated with the prevalence or decline, 
and consequent future disappearance, 
of the infection in a host population. 
The system evolution to any of these scenarios is determined by the basic reproduction ratio, 
which, 
for the normalized SEIRS model according to Eqs.~(\ref{eq_seirNormalizado}), 
is given by 
\begin{equation}
	\mathcal{R}_0 = \alpha \beta(t)/\left[(\alpha + \eta)(\gamma + \eta)\right],   
	\label{eq_R0}
\end{equation}
in the same way as the autonomous SEIRS~\cite{Rock2014,Bjornstad2020}, 
but including an explicit time dependency in $\beta(t)$. 
$\mathcal{R}_0$ oscillates periodically in the range 
\begin{equation}
	\mathcal{R}_0^{-} \leq \mathcal{R}_0 \leq \mathcal{R}_0^{+}, 
	\label{eq_intervaloR0}
\end{equation}
being the edge values defined as 
\begin{equation}
	\mathcal{R}_0^{\pm} \coloneqq \frac{\alpha \beta_0(1\pm\beta_1)}{(\alpha + \eta)(\gamma + \eta)}. 
	\label{eq_definicaoLimitesR0}
\end{equation}

We said that $\mathcal{R}_0$ is strictly greater than unity when $\mathcal{R}_0^{+}\geq\mathcal{R}_0^{-}>1$ and, 
on the other hand, 
refer it to by strictly less than unity if $\mathcal{R}_0^{-}\leq\mathcal{R}_0^{+}<1$. 
For autonomous compartmental epidemiological models, 
it is known that $\mathcal{R}_0<1$ leads to eradication of infection~\cite{Bjornstad2018}. 
However, 
since the system is non-autonomous, 
specifically with the transmission rate given as a function of time, 
a sufficient condition for convergence to the DFE solution is the long-term basic reproduction ratio $\overline{\mathcal{R}}_0<1$, 
which is calculated over the long-term average of the system evolution~\cite{Seir_Sazonal}.  
{Also, 
	the point $P_{\rm DFE}$ is asymptotically stable if $\mathcal{R}_0^{+}<1$, 
	i.e., 
	verifying the parameters relation $\beta_0(1 + \beta_1) < (\alpha + \eta)(\gamma + \eta)/\alpha$.}
{
\subsection{Around endemic equilibrium}
The autonomous SEIRS model, 
with $\beta \equiv \beta_0$ (equivalent to $\beta_1 = 0$ in Eq.~(\ref{eq_betaPeriodica})), 
presents a second equilibrium state, 
denoted by endemic equilibrium point $P_{\rm EE}(s_{\rm e}, e_{\rm e}, i_{\rm e})$, 
in which there is no extinguishes the infection in the host population~\cite{Rock2014}. 
In the non-autonomous case addressed here, 
the trajectories can oscillate around this point. 
We illustrate this behavior in Fig.~\ref{fig_EquilibrioEndemico}. 
Arises from the fixed point condition,  
we obtain  
\begin{equation}
	s_{\rm e} = \frac{1}{\mathcal{R}_{\rm c}};~~
	e_{\rm e} = \frac{\gamma + \eta}{\alpha\beta_0} \varepsilon;~~
	i_{\rm e} = \frac{\varepsilon}{\beta_0}, 
\end{equation}
where $\mathcal{R}_{\rm c}$ is the constant basic reproduction ratio, 
with $\beta \equiv \beta_0$ in Eq.~(\ref{eq_R0}), 
and 
\begin{equation}
	\varepsilon \coloneqq \frac{(\delta + \eta)(\mathcal{R}_{\rm c} - 1)}{1 + \frac{\mathcal{R}_{\rm c} \delta}{\beta_0} \left[ 1 + \frac{\gamma + \eta}{\alpha} \right]}.
\end{equation}

Note that, 
to recover a SEIR scheme from the system~(\ref{eq_seirNormalizado}), 
simply set $\delta = 0$, 
resulting in the $P_{\rm EE}$ for a scenario with permanent immunity~\cite{aron1984}. 
Considering the autonomous model, 
it is necessary $\mathcal{R}_{\rm c}>0$ for $P_{\rm EE}$ to be attractive~\cite{aron1984,schwartz1983}. 
To exemplify the convergence towards endemic equilibrium, 
we numerically integrate the autonomous system with the follow parameters: 
$\beta_0=280$, 
$\alpha=\gamma=100$, 
$\delta=2.5\times10^{-1}$ and 
$ \eta = 2\times10^{-2}$. 
In this configuration, 
$\mathcal{R}_{\rm c} \approx 2.799$, 
$s_{\rm e} \approx 3.573\times10^{-1}$ and 
$i_{\rm e} \approx 1.735\times10^{-1}$. 
Figure~\ref{fig_EquilibrioEndemico} shows the projection of the system's trajectories onto the $i\times s$ plane, 
with $P_{\rm EE}$ in red  
\begin{figure}[!t]
	\centering
	\includegraphics[width=0.92\columnwidth]{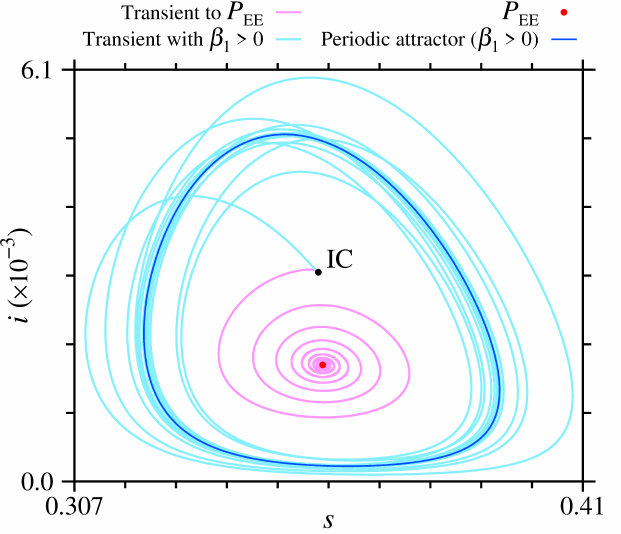}
	\caption{{Trajectories evolving from ${\rm IC}$ to the attractors in the autonomous and 
		non-autonomous cases: 
		pink spiral converging at $P_{\rm EE}$ (red dot) in the autonomous case; 
		light blue curve converging to the periodic attractor (blue line) in the non-autonomous case. 
		We use a fourth order Runge-Kutta numerical integration method with a fixed step of $10^{-3}$.  
		Transients smaller than $5\times10^5$ integration steps being sufficient for the trajectories convergence to the attractors with great accuracy.}} 
	\label{fig_EquilibrioEndemico}
\end{figure}
and a periodic orbit (blue line), 
which results from the evolution of the non-autonomous model with $\beta_1 = 0.1$ and 
$\omega = 2\pi$. 
Pink and light blue curves are the transient trajectories spiraling from the initial condition ${\rm IC(0.3564, 0.0032, 0.0031)}$ 
(highlighted point) to $P_{\rm EE}$ and 
the periodic attractor, 
respectively. 

%Given the similarity between the two models, 
%with and 
%without reinfection, 
%we adopt the same sequence of procedures performed by Schwartz and 
%Smith in their work on subharmonic bifurcations in an SEIR model with periodic contact rate~\cite{schwartz1983}. 
%We consider a coordinate transformation, 
%where $P_{\rm EE}$ becomes the origin, 
%as follows 
%\begin{equation}
%	s = s_{\rm e}(1 + x);~~
%	e = e_{\rm e}(1 + y);~~
%	i = i_{\rm e}(1 + z).
%	\label{eq_mudancaVariavelPEE}
%\end{equation}
%In order to obtain algebraic simplifications, 
%we use three auxiliary parameters, 
%$C_{\delta},C_{\alpha},C_{\gamma}>0$, 
%such that 
%\begin{equation}
%	(\delta + \eta) = \frac{C_{\delta}}{\varepsilon};~~
%	(\alpha + \eta) = \frac{C_{\alpha}}{\varepsilon};~~
%	(\gamma + \eta) = \frac{C_{\gamma}}{\varepsilon}.
%	\label{eq_constantesAuxiliares}  
%\end{equation}
%Substituting (\ref{eq_mudancaVariavelPEE}) and 
%(\ref{eq_constantesAuxiliares}) in Eqs.~(\ref{eq_seirNormalizado}),  
%we obtain 
%\begin{align}
%	\frac{dx}{dt} &= -\varepsilon \left[ \frac{\beta (t)}{\beta_0}(1 + x)(1 + z) - 1 \right] + \nonumber \\
%	&-\varepsilon \left[ \frac{\mathcal{R}_{\rm c}\delta }{\beta_0} \left( \frac{C_{\gamma}}{\varepsilon \alpha} y + z \right)\right]
%	- \frac{C_{\delta}}{\varepsilon}x,\\
%e_{\rm e} &= \frac{\gamma + \eta}{\alpha} i_*,\\
%i_{\rm e} &= \frac{\alpha}{\gamma + \eta}e_*.
%\end{align}
}
%\begin{equation}
%	\beta_0(1 + \beta_1) < \frac{(\alpha + \eta)(\gamma + \eta)}{\alpha}. 
%\end{equation}}

%{An alternative to compute $\mathcal{R}_0$ for a periodic time dependent transmission rate is taking the average~\cite{Wesley2009,Greenhalgh2003} 
%	\begin{equation}
%		\langle{\beta (t)}\rangle = \frac{1}{T} \int_{0}^{T} \beta (t) dt.
%		\label{beta_medio}
%	\end{equation}
%	Where $T$ is the periodo of the function $\beta(t)$. 
%	Considering $\langle{\beta (t)}\rangle$ the stability analysis remains valid~\cite{Wesley2009}.} 
%{We get the basic reproduction number as a function of this average:
%	\begin{equation}
%		\overline{\mathcal{R}}_0 \coloneqq \frac{\alpha \langle{\beta (t)}\rangle}{(\alpha + \eta)(\gamma + \eta)}. 
%	\end{equation}
%	}
%
%==================================================================================================
%
\section{NUMERICAL RESULTS AND DISCUSSION} \label{sec_resultados} 
In this section, 
we numerically investigate the SEIRS model dynamics emphasizing the influence of seasonality, 
included according to Eq.~(\ref{eq_betaPeriodica}). 
To evolve the system~(\ref{eq_seirNormalizado}), 
we employ the fourth-order Runge-Kutta integration method with a fixed time step of $10^{-3}$ and 
consider $5\times10^5$ integration steps as transient. 
{Our simulations indicate that these settings are sufficient for the convergence of the trajectory and 
	tangent vectors~\cite{benettin1980,Shimada1979}.}
{In particular, 
	we check the minimum transient required in parameter planes with grids of $200\times200$ points, 
	at $8$ transient values from $10^3$ to $5\times10^6$ integration steps. 
	We found that $5\times10^5$ steps are enough. 
	We also check this quantity for the other results.} 
{The time unit in the simulation is one year and the parameters have unit $\text{year}^{-1}$, 
	except $\beta_1$ which is dimensionless and $[\omega]\equiv \text{rad/year}$.} 
{We vary the parameters over wide intervals, 
	aiming to cover characteristic values of several infections,  
	 for example, 
	 measles ($\alpha=38.5$ and $\gamma=100$ )~\cite{Bai2012}, 
	 influenza ($\alpha=228.12$ and $\gamma=125$)~\cite{cori2012}, 
	 and others~\cite {lessler2009, sartwell1950}.}
We start by analyzing the Lyapunov exponents on the parameter planes shown in Sub-s.~\ref{sec_lyapunov}. 
Next, 
in Sub-s.~\ref{sec_multiestabilidade}, 
we highlight and study the system multistability. 

\subsection{Lyapunov exponents analysis} \label{sec_lyapunov}
In the present study, 
the system dynamics characterization %was carried out mainly through the largest Lyapunov exponent. 
is made mainly by the largest Lyapunov exponent. 
We verify that the SEIRS model, 
under the influence of a seasonal transmission rate, 
can evolve both to chaotic and 
periodic trajectories, 
depending on the parameters and
initial conditions (see Sec.~\ref{sec_multiestabilidade}). 
%for many configurations, 
%also on the initial conditions adopted, 
%according to the results shown in Sec.~\ref{sec_multiestabilidade}. 
%To examine the relevance of seasonality in the system dynamics, 
%more precisely, 
%to investigate the effects arising from its frequency $\omega$, 
%whose associated period is $T = 2\pi/\omega$, 
To investigate the effects of seasonality frequency $\omega$, 
with period $T = 2\pi/\omega$, 
in the system dynamics, 
we compute the Lyapunov spectrum along parameter planes $[\cdot] \times \omega$, 
where $[\cdot]$ is a model parameter. 
Previously,
we rewrite the system~(\ref{eq_seirNormalizado}) in autonomous form,  
thus we perform the transformation $\omega t \mapsto \tau \in [0,2\pi)$ and 
include the respective differential equation $d\tau/dt = \omega$. 
The Lyapunov spectrum is obtain 
{evolving the system and its respective linearized equations} 
using the algorithm described by Wolf~\textit{et al}~\cite{Wolf1985} and 
the exponents sorted in descending order: 
$\lambda_1 \geq \lambda_2 \geq \lambda_3 \geq \lambda_4$.   
We adopt the initial condition $(s_0,e_0,i_0,\tau_0) = (0,999,0,0,001,0)$ and 
consider $10^6$ integration steps, 
after discarding the transient. 
Without prejudice, 
throughout the text %we avoid the information 
$\tau_0=0$ will be omitted. 
%bearing in mind that this value refers to the initial time. 

\begin{figure}[!b]
	\centering
	\includegraphics[width=0.92\columnwidth]{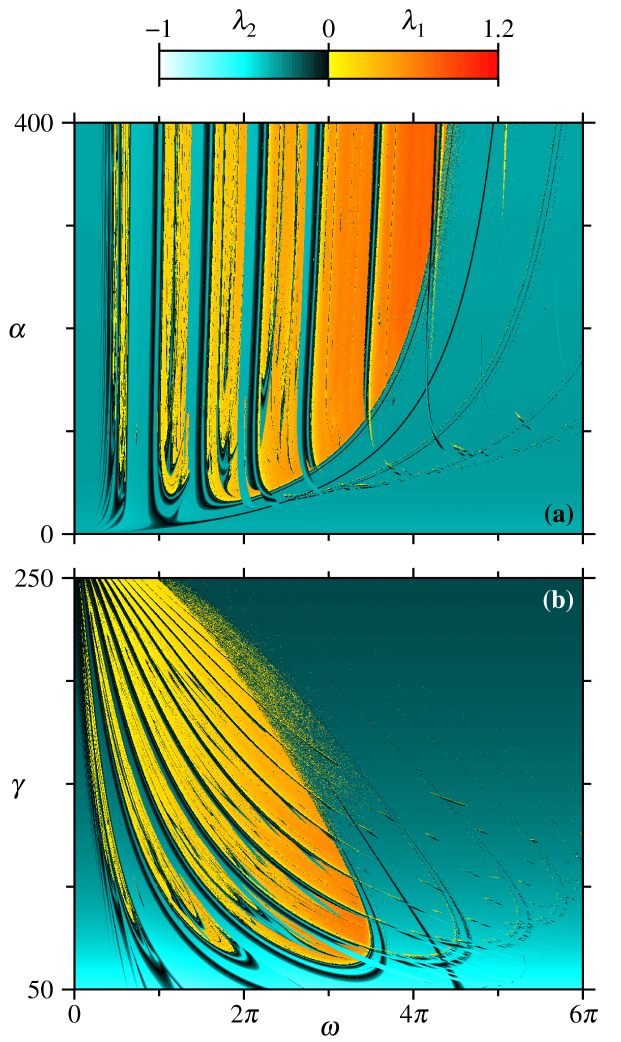}
	\caption{Parameter planes discretized in uniform grids of $1000\times1000$ points. 
		Lyapunov exponents in color, 
		according to legend. 
		Where $\lambda_1=0$, 
		we display $\lambda_2$. 
		Chaotic regions in the gradient from yellow to red, 
		periodic ones in cyan shades. 
		Horizontal axis shows the seasonality frequency. 
		(a) Inverse of latent period on vertical axis. 
		Intervals of chaotic regions in bands approximately parallel to the $\alpha$ axis reveal the greater relevance of $\omega$ in the system evolution. 
		(b) Recovery rate on the vertical axis. 
		The arrangement of the chaotic bands shows $\gamma$ as a determining factor for the dynamics. 
		Shrimp-like periodic structures close to $\alpha=\gamma=100$ for the frequences $\omega\approx\pi$ and 
		$\omega\approx2\pi$.} 
	\label{fig_alfaGamaOmega}
\end{figure}

\begin{figure}[!b]
	\centering
	\includegraphics[width=0.92\columnwidth]{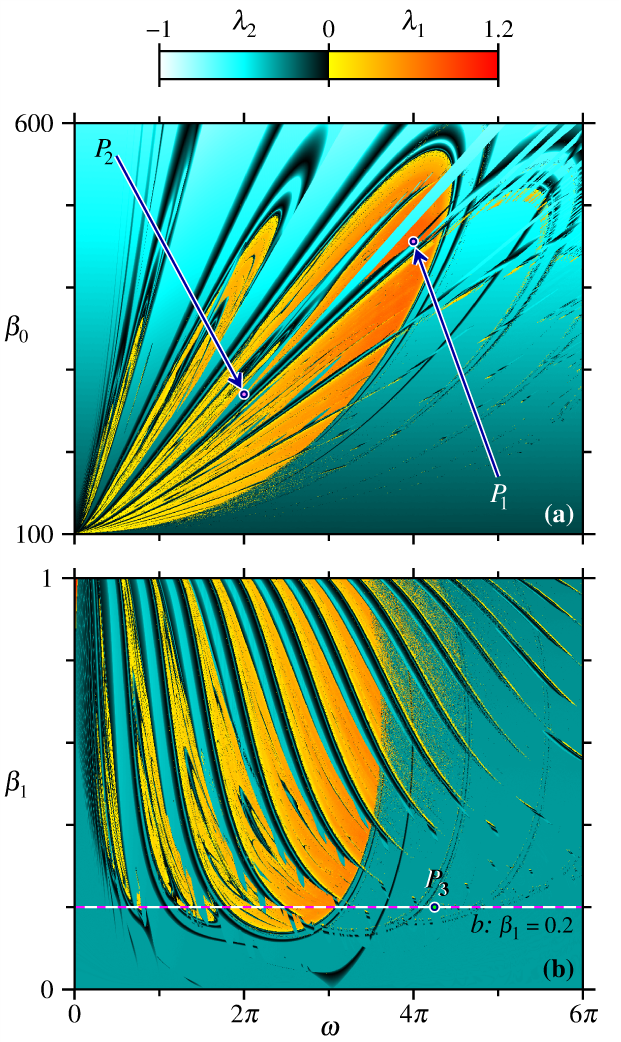}
	\caption{Parameter planes discretized in uniform grids of $1000\times1000$ points. 
		Lyapunov exponents in color, 
		according to legend. 
		Where $\lambda_1=0$, 
		we display $\lambda_2$. 
		Chaotic regions in the gradient from yellow to red, 
		periodic ones in cyan shades. 
		Horizontal axis shows the seasonality frequency. 
		(a) Mean transmission rate on vertical axis. 
		Diagonal periodic bands reveal the joint relevance of $\beta_0$ and 
		$\omega$ to the system evolution. 
		Highlighted points $P_1(4\pi,456)$ and 
		$P_2(2\pi,270)$ analyzed in Sub-s.~\ref{sec_multiestabilidade} Figs.~\ref{fig_atratorP1} and 
		\ref{fig_baciaP2}. 
		(b) seazonality degree on vertical axis. 
		Emphasized the segment $b:~\beta_1=0.2,~\text{whith}~0<\omega\leq6\pi$ (dashed line) and 
		the point $P_3(4.25\pi,0.2)$, 
		for which we analyze bifurcation diagrams and 
		attraction basins, 
		respectively (see Fig.~\ref{fig_bifurcacaoBaciaP3}).} 
	\label{fig_beta0Beta1Omega}
\end{figure}

%Here, 
We analyze four parameter planes calculated on uniform grids of $1000\times1000$ points, 
displayed in Figs.~\ref{fig_alfaGamaOmega} and \ref{fig_beta0Beta1Omega}. 
Transmissivity oscillation frequency is on the horizontal axis, 
which is evaluated in the range $0<\omega\leq6\pi$. 
The two largest Lyapunov exponents are represented in color: 
chaotic regions with $0 < \lambda_1 \leq 1.2$ (gradient from yellow to red); 
in periodic ones ($\lambda_1 = 0$) it shows $-1 \leq \lambda_2 < 0$, 
starting from white ($\lambda_2 = -1$), 
passing through shades of cyan to black color ($\lambda_2 = 0$). 

Figure~\ref{fig_alfaGamaOmega} illustrates two planes formed by combining $\omega$ with typical epidemiological model parameters. 
In the panel Fig.~\ref{fig_alfaGamaOmega}(a) we analyze the interval $0<\alpha\leq400$ on the vertical axis, 
being fixed: 
$\gamma=100$, 
$\eta=0.02$, 
$\delta=0.25$, 
$\beta_0 = 270$ and 
$\beta_1 = 0.28$. 
With this parameter setting $0<\mathcal{R}_0\leq3.455$, 
such that for $\alpha<0.008$ the basic reproduction ratio is strictly less than unity. 
In the interval $0.008<\alpha<0.021$ we have $0.563<\mathcal{R}_0^{-}<1$ and 
$\mathcal{R}_0^{+}>1$. 
Complementarily, 
we obtain $1<\mathcal{R}_0~\forall~t$ with $0.021<\alpha$ and, 
given the $\alpha$ axis discretization in steps of $\Delta \alpha=0.4$, 
in Fig.~\ref{fig_alfaGamaOmega}(a) we only observe results with the basic reproduction ratio strictly greater than unity. 
The evolution of the system to chaotic dynamics is predominantly determined by the seasonality frequency. 
Seasonal cycles of less than $6$ months lead predominantly to periodic behavior, 
see the wide dark cyan band from $\omega=4\pi$, 
with small chaotic regions inserted there. 
For $\omega<4\pi$ approximately vertical bands occur,  
revealing that latency intervals smaller than $\approx 1.3$ days,
corresponding to $\alpha > 280$, 
are of little relevance to the dynamics. 
Next to $\alpha=100$ and 
$\omega=2\pi$, 
there is a shrimp-like periodic structure, 
also seen around $\omega\approx\pi$. 
Such parameter values correspond to the latency $\approx3.6$ days and 
seasonality with periods $T=1$ and $T=2$ years, 
respectively. 
{Shrimps recurrently appear in parameter planes of paradigmatic non-linear systems~\cite{camarao_2,camarao_3,camarao_4,rech2017embed} 
	from a peculiar arrangement of two saddle-node bifurcation curves and 
	routes to chaos via period doubling~\cite{gallas1994dissecting,camarao5}.}
{The presence of these structures is evidence of rich dynamics and} 
this vicinity is known to display shrimp cascades forming a repeating pattern with self-similarity~\cite{gallas1994dissecting,Gallas1995}. 
%In these regions, 
%the system dynamics is susceptible to very small variations in parameter values. 

In Fig.~\ref{fig_alfaGamaOmega}(b) we show the plane $\gamma\times \omega$, 
with $50<\gamma\leq250$, 
$\alpha=100$ and 
the other parameters are kept equal to those used in panel (a). 
For this system configuration, 
we calculate $\mathcal{R}_0>1~\forall~t$ with $\gamma<194.4$. 
However, 
in the range $194.4<\gamma\leq250$ we have $0.777<\mathcal{R}_0^{-}<1$ and 
$\mathcal{R}_0^{+}>1$.
Shrimp-like periodic structures are observed in the vicinity of $\omega=\pi$ and 
$\omega=2\pi$, 
with $\gamma\approx100$. 
These periodic regions embedded in the chaotic bands are related to those in the $\alpha \times \omega$ plane. 
For the adopted parameters, 
it can be observed that the dynamics is more influenced by the infectious period than the latent one. 
For lower recovery rates, 
in the range $50<\gamma\leq100$ corresponding to infectious periods between $\approx7.3$ and 
$3.6$ days, 
chaotic bands extend to $\omega\approx3.5\pi$. 
While for periods smaller than $\approx1.5$ days, 
referring to $\gamma\approx250$, 
the chaotic bands compress into the interval $0<\omega\leq\pi$. 
Thus, 
for small infectious periods, 
chaotic trajectories occur only for seasonality with $T\geq2$ years. 
Similar to what is observed in panel (a), 
from $\omega\approx 3.5\pi$ ($T<6.9$ months) 
the analyzed parameters planes presents a wide range of periodic dynamics. 
Cyan area taken in the intervals $50<\gamma<75$ and 
$0<\omega\leq6\pi$ corresponds to more stable periodic orbits, 
{i.e. these are less sensitive to small disturbances} 
than those obtained for higher recovery rates, 
notably the dark cyan region with $\gamma \rightarrow 250$. 

As seen, 
the frequency of seasonal cycles significantly affects the SEIRS model dynamics, 
in addition to this factor, 
we highlight the effects of the mean transmission rate and 
the seasonality degree. 
Figure~\ref{fig_beta0Beta1Omega} display two planes with these parameters on the vertical axes, 
we set $\alpha=\gamma=100$ and 
the other values equal to those used in Fig.~\ref{fig_alfaGamaOmega}, 
being in panel (a) $\beta_1 =0.28$ and 
in (b) $\beta_0 =270$. 
Figure~\ref{fig_beta0Beta1Omega}(a) comprises the interval $100<\beta_0\leq600$ on the vertical axis, 
in it we see a periodic region around $\omega=2\pi$ and 
$\beta_0=300$ immersed in a chaotic band. 
This shrimp-like structure corresponds to that shown in Fig.~\ref{fig_alfaGamaOmega}(b). 
Mean transmission rate increasing ($\beta_0\rightarrow600$)  
is associated with the occurrence of very stable periodic orbits ($\lambda_2<-0.5$). 
%compared to those obtained along the other parameter planes. 
On the other hand, 
also for large transmissivities, 
with $300<\beta_0<550$ and 
around $\omega=4\pi$, 
the chaotic trajectories occur with the highest values of $\lambda_1$ (colors from orange to red). 
In the $\beta_0\times\omega$ plane evaluated section, 
the periodic bands are interspersed diagonally with the chaotic ones, 
such that the system evolution depends both on the mean transmission rate and 
on its frequency. 
In this range of $\beta_0$ the basic reproduction ratio varies between $\mathcal{R}_0^{-}\approx0.720$ and 
$\mathcal{R}_0^{+}\approx7.677$. 
For values of $\beta_0>138.944$ results $\mathcal{R}_0>1~\forall~t$. 
In the interval $100<\beta_0<138.944$ we get $0.720<\mathcal{R}_0^{-}<1$ and 
$\mathcal{R}_0^{+}>1$. 
Points $P_1(4\pi,456)$ and 
$P_2(2\pi,270)$, 
highlighted in Fig.~\ref{fig_beta0Beta1Omega}(a), 
are targets of the multistability analysis shown in Sub-s.~\ref{sec_multiestabilidade}. 

Figure~\ref{fig_beta0Beta1Omega}(b) illustrates the influence of $\beta_1$ in the SEIRS model dynamics. 
We consider the interval $0<\beta_1\leq1$ on the vertical axis, 
where the transmission rate oscillates from the minimum $\beta_0(1-\beta_1)$ to the maximum $\beta_0(1+\beta_1)$ within one period $T$. 
Here $0<\mathcal{R}_0<5.398$, 
where $\mathcal{R}_0$ is strictly greater than unity for $0<\beta_1<0.629$. 
In the seasonality degree range $0.629<\beta_1\leq1$, 
the edge values are $0\leq\mathcal{R}_0^{-}<1$ and 
$\mathcal{R}_0^{+}>1$. 
We observe a pattern of chaotic bands interspersed with periodic ones, 
similar to that displayed in Fig.~\ref{fig_alfaGamaOmega}(a). 
However, 
$\beta_1$ is more relevant to the dynamics than the parameter $\alpha$, 
increasing its influence to higher frequencies of seasonal cycles from $\omega\approx\pi$. 
In Sub-s.~\ref{sec_multiestabilidade} we evidence the system multistability along the segment highlighted in $\beta_1=0.2$ 
(magenta and white dashed line) and 
for the point $P_3(4.25\pi,0.2)$. 
\subsection{Multistability and coexistence of chaotic and periodic attractors} \label{sec_multiestabilidade} 
System~(\ref{eq_seirNormalizado}) with periodic $\beta\equiv\beta(t)$ %, 
%according Eq.~(\ref{eq_betaPeriodica}), 
presents multistability~\cite{Gabrick2023}, 
i.e., 
different orbits can occur for a given parametric configuration, 
depending on the initial condition~\cite{Feudel1997}. 
Given this feature, 
in addition to the unpredictability due to chaotic trajectories, 
the coexistence of chaotic and 
periodic attractors is observed, 
as well as distinct periodic orbits. 
Additionally to the varied dynamic behaviors, 
slightly different starting conditions can lead to more pronounced peaks in the infectious curve. 
In this subsection we investigate multistability in the model, 
especially for the parameter values highlighted in Sub-s~\ref{sec_lyapunov}. 
All basins of attraction shown below employ a uniform grid plane discretization of $1000\times 1000$ points. 
\vspace{0.3cm} 

\begin{figure}[!h]
	\centering
	\includegraphics[width=0.92\columnwidth]{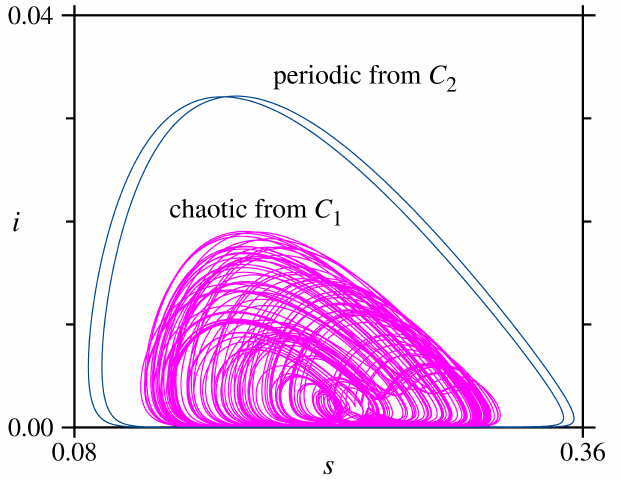}
	\caption{Plane $i\times s$ projection of attractors obtained for the point $P_1(4\pi,456)$,
		highlighted in Fig.~\ref{fig_beta0Beta1Omega}(a). 
		Chaotic attractor (magenta line) obtained from the initial condition $C_1 = (0,990,0,001,0,009)$ and 
		periodic one (blue line) resulting from the initial condition $C_2 = (0.998,0.001,0.001)$. 
		For the chaotic case the maximum value of infectious agents is $i_{\rm max}\approx0.02$, 
		in the periodic one it is $i_{\rm max}\approx0.032$.} 
	\label{fig_atratorP1}
\end{figure}
\vspace{0.1cm}

Figure~\ref{fig_atratorP1} exhibits two attractors projected in the plane $i \times s$, 
one chaotic (magenta line) and 
other periodic (blue line), 
both at the point $P_1(\omega,\beta_0)=P_1(4\pi,456)$ shown in Fig.~\ref{fig_beta0Beta1Omega}(a). 
%in which the other parameters assume the values: 
%$\eta=0.02$, 
%$\delta=0.25$, 
%$\beta_1=0.28$ and 
%$\alpha=\gamma=100$. 
The chaotic attractor is generated from the initial condition $C_1 = (0.990,0.001,0.009)$ and 
shows the maximum infectious population proportion $i_{\rm max}\approx 0.02$. 
For the periodic one we adopt the initial condition $C_2 = (0.998,0.001,0.001)$, 
resulting the maximum value $i_{\rm max}\approx 0.032$. 
The periodic case recurrently leads to peaks of infections $\approx 50\%$ greater than the maximum observed in the chaotic situation. 
If, 
on the one hand, 
predictability facilitates the planning of epidemic containment protocols, 
on the other hand, the greater number of cases can cause harm to public health. 
However, 
both chaotic and periodic orbit present the same time average of cases $\langle{i}\rangle_t \approx 2.1\times 10^{-3}$

\begin{figure}[!t]
	\centering
	\includegraphics[width=0.92\columnwidth]{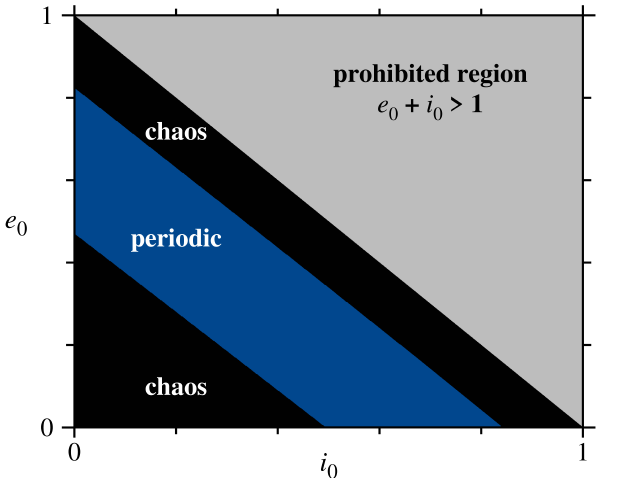}
	\caption{Plane $e_0\times i_0$ of initial conditions discretized in uniform grid of $1000\times 1000$ points. 
		Parametric configuration of point $P_2(2\pi,270)$. 
		In gray we identify the region outside the model domain.
		Of all evaluated valid initial conditions, 
		$\approx46.5\%$ evolve to periodic orbits (blue bands),
		the remaining $\approx53.5\%$ leads to chaotic ones (black color).} 
	\label{fig_baciaP2}
\end{figure}

For the point $P_2(\omega,\beta_0)=P_2(2\pi,270)$, 
also displayed in Fig.~\ref{fig_beta0Beta1Omega}(a), 
we perform a scan of initial conditions and 
distinguish the obtained orbits between chaotic and 
periodic ones. 
To that end, 
we uniformly vary the initial values of infectious ($i_0$) and 
exposed ($e_0$). 
Figure~\ref{fig_baciaP2} illustrates the plane $e_0\times i_0$ %discretized in uniform grid of $1000\times 1000$ points, 
with $0<e_0,i_0\leq1$ and 
$s_0 = 1 - (e_0 + i_0)$. 
Basins are identified in colors, 
where pairs $(i_0,e_0)$ that lead to chaotic attractors are in black and 
those that lead to periodic behavior are in blue. 
%The gray region violates the constraint between the variables. 
The gray region is outside the model domain. 
%due to the constraint between the variables.   
%In this evaluation, 
We find $\approx53.5\%$ of valid initial conditions (outside the gray region) leading to chaotic behavior. 
Pairs $(e_0,i_0)$ that evolve to periodic attractors have a sum $0.471\leq e_0 + i_0\leq0.824$ and, 
consequently, 
$0.176\leq s_0\leq0.529$. 
These proportions of infectious and exposed individuals are very high for the initial phase of an epidemic, 
even so, 
when it comes to dynamic analysis, 
these data help to understand the basins of each behavior and 
are especially valuable for the development of epidemic control protocols. 

\FloatBarrier

\begin{figure}[!t]
	\centering
	\includegraphics[width=0.92\columnwidth]{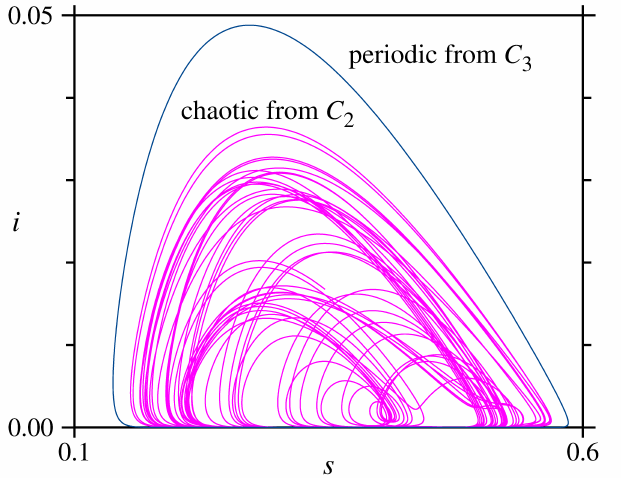}
	\caption{Plane $i\times s$ projection of attractors obtained for the point $P_2$.
		%highlighted in Fig.~\ref{fig_beta0Beta1Omega}(a). 
		Chaotic attractor (magenta line) results from the initial condition $C_2 = (0.998,0.001,0.001)$, 
		periodic one (blue line) arising from $C_3 = (0.399,0.001,0.6)$. 
		The chaotic presents a maximum of infectious $i_{\rm max}\approx0.037$ and  
		the periodic one has $i_{\rm max}\approx0.048$.} 
	\label{fig_atratorP2}
\end{figure}

Similar to what is verified for the point $P_1$, 
for $P_2$ we find a periodic orbit with a higher peak of infectious agents than the chaotic one, 
as shown in Fig.~\ref{fig_atratorP2}. 
The chaotic attractor (magenta) is obtained from the same initial condition $C_2$ adopted in Fig.~\ref{fig_atratorP1}, 
the periodic one (blue) results from the condition $C_3 = (0.399,0.001,0.6)$. 
Maximum infectious cases in the chaotic scenario is $i_{\rm max}\approx0.037$, 
although in the periodic trajectory we have $i_{\rm max}\approx0.048$. 
Both scenarios present approximately the same time average of cases,  
being $\langle{i}\rangle_t \approx 1.7\times 10^{-3}$. 
\FloatBarrier

\begin{figure}[!b]
	\centering
	\includegraphics[width=0.92\columnwidth]{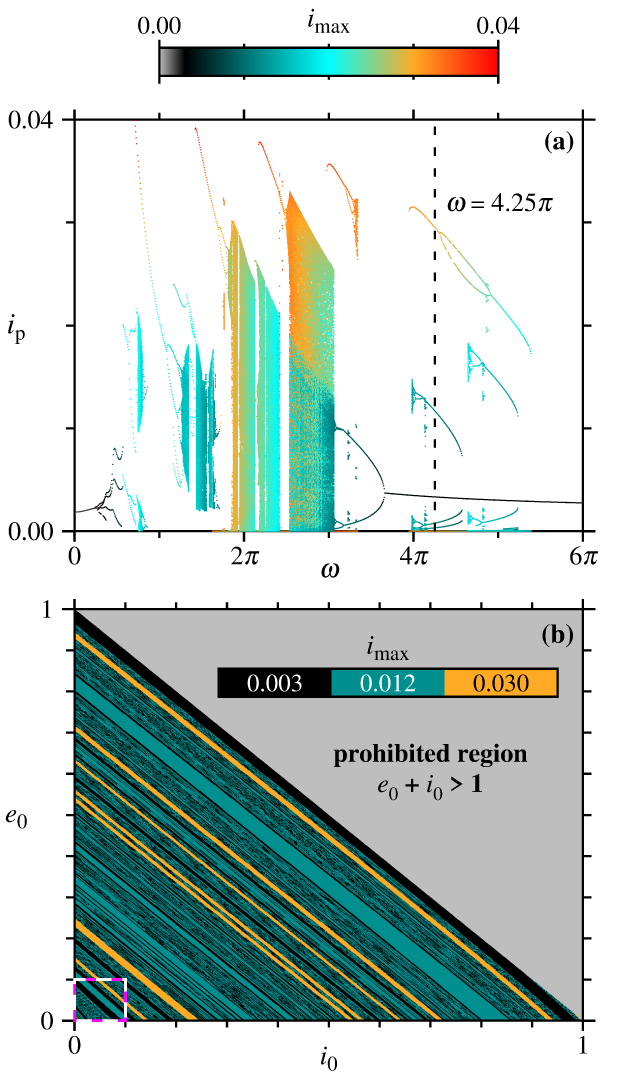}
	\caption{(a) Concatenated bifurcation diagrams of $100$ initial conditions for each $\omega$ value along segment $b_1$, 
		dashed line in Fig.\ref{fig_beta0Beta1Omega}(b). 
		Horizontal axis discretized into $1000$ equidistant points. 
		Plotted all distinct $\mathcal{I}_k(\omega)$. 
		Color code for $i_{\rm max}(\omega;C_k)$ distinguishing attractors, 
		according Eq.~(\ref{eq_maximoInfeccioso}). 
		For $\omega=4.25\pi$ (dashed line) there are $3$ periodic attractors, 
		this point corresponds to $P_3(4.25\pi,0.2)$ in Fig.\ref{fig_beta0Beta1Omega}(b). 
		(b) Basins of attraction in the initial conditions plane $e_0\times i_0$ discretized in uniforme grid of $1000\times 1000$ points, 
		with $s_0=1 - (e_0 + i_0)$ and 
		parametric configuration for $P_3$. 
		The gray region is outside of the normalized system domain. 
		Dashed square in the bottom left corner is magnified in Fig.~\ref{fig_baciaP5zoom}(a).} 
	\label{fig_bifurcacaoBaciaP3}
\end{figure}

Figure~\ref{fig_bifurcacaoBaciaP3}(a) concatenates $100$ bifurcation diagrams along segment $b:\beta_1=0.2$, 
in the frequency interval $0<\omega\leq6\pi$ uniformly discretized in $1000$ points,  
see dashed line in Fig.\ref{fig_beta0Beta1Omega}(b). 
For each value of $\omega$, 
$100$ trajectories are generated from randomly assigned initial conditions in the interval $0<e_0,i_0<1$, 
drawn in a uniform distribution and respecting the restriction $e_0 + i_0 \leq 1$, 
being $s_0 = 1 - (e_0 + i_0)$. 
Once the transient has been discarded, 
we continue to evolve the system for $10^6$ integration steps and 
select the local maxima $i_{\rm p}$ (peak value) 
in the infectious time series over the last $7.5\times10^4$ steps, 
equivalent to the last $75$ years in simulation. 
In this way, 
we construct a set $\mathcal{I}_k(\omega)$ of the local maxima in $i(t)$ curves, 
for given $\omega$ and 
the $k$-th initial condition. 
If there are periodic attractors, 
it can occur $\mathcal{I}_j(\omega)=\mathcal{I}_{k\neq j}(\omega)$, 
reducing the total amount of sets. 
Given the $k$-th inicial condition $C_k$, 
we have 
\begin{equation}
	i_{\rm p}(\omega;C_k) \coloneqq i(t_{\rm p},\omega;C_k), 
\end{equation}
being 
\begin{equation}
	\frac{di}{dt}\bigg|_{t_{\rm p}} = 0~\text{and}~~\frac{d^2i}{dt^2}\bigg|_{t_{\rm p}} < 0. 
\end{equation}
Then, 
we define these sets as 
\begin{align}
	\mathcal{I}_k(\omega) \coloneqq \left\{i_{\rm p}(\omega;C_k) : t_{\rm p}\in\mathcal{T}\right\}, 
	\label{eq_conjuntoI}
\end{align}
where $\mathcal{T}$ is the evaluation time interval. 
Complementarily, 
we identify each different set through its maximum value 
\begin{equation}
	i_{\rm max}(\omega;C_k) \coloneqq \max\{\mathcal{I}_k(\omega)\}.
	\label{eq_maximoInfeccioso}
\end{equation}

In Fig~\ref{fig_bifurcacaoBaciaP3}(a) we plot all the distinct $\mathcal{I}_k(\omega)$ sets. 
By means of $i_{\rm max}$ it is possible to distinguish different orbits that occur for the same parametric configuration, 
%as shown in Fig.~\ref{fig_bifurcacaoBaciaP3}(a), 
where we use the color code for such identification. 
This distinction is effective between periodic orbits and 
is also useful for highlighting points corresponding to smaller chaotic attractors from larger ones. 
Figure~\ref{fig_bifurcacaoBaciaP3}(a) displays the coexistence of distinct chaotic orbits in the band starting at $\omega\approx2.5\pi$ and 
goes to $\omega\approx3\pi$, 
where chaotic attractors occur with $i_{\rm max}\approx0.030$ (orange dots) and 
others with $i_{\rm max}\approx0.020$ (cyan and teal dots), 
depending on the initial conditions. 
For higher frequencies,
we observe a vast coexistence interval of periodic attractors, 
as an example in $\omega=4.25\pi$ (vertical dashed line), 
where are $3$ periodic orbits, 
being these of: 
period $1$, with $i_{\rm max}\approx0.003$ (black color); 
period $3$, with $i_{\rm max}\approx0.012$ (teal color); and 
period $2$, with $i_{\rm max}\approx0.030$ (orange color). 
High peaks of infectious are obtained for $\omega\approx1.5\pi$ and 
$\omega\approx0.7\pi$, 
where $i_{\rm p}\approx0.04$, 
i.e., 
the local maxima in the infectious time series reach, 
periodically, 
nearly $4\%$ of the host population.

\begin{figure}[!t]
	\centering
	\includegraphics[width=0.92\columnwidth]{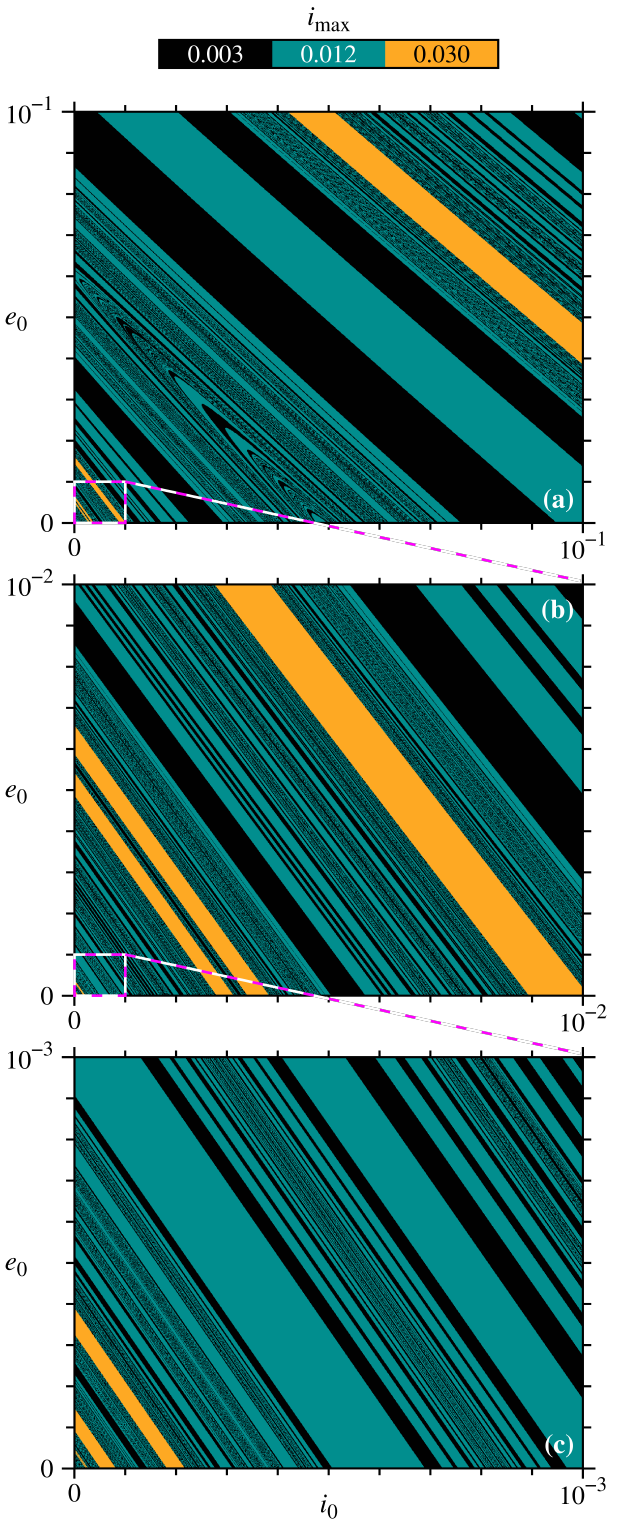}
	\caption{Attraction basins in a succession of magnifications. 
		Axes discretized in uniform grid of $1000\times 1000$ points. 
		Color code identifies the maximum infectious value $i_{\rm max}$ of each attractor, 
		as defined in Eq.~(\ref{eq_maximoInfeccioso}). 
		(a) Magnification of the delimited region in Fig.~\ref{fig_bifurcacaoBaciaP3}(b), 
		with $0<e_0,i_0\leq10^{-1}$. 
		Box in bottom left corner enlarged on panel 
		(b) $0<e_0,i_0\leq10^{-2}$. 
		In turn, the highlighted box is magnified in  
		(c) $0<e_0,i_0\leq10^{-3}$.} 
	\label{fig_baciaP5zoom}
\end{figure}

Coexistence of orbits with such a discrepancy in the maximum value of infectious, 
as seen in Fig.~\ref{fig_bifurcacaoBaciaP3}(a), 
evidencing the initial conditions influence,  
not only to the system dynamic regime, 
but also determining the system evolution to scenarios with a greater or lesser number of infected individuals, 
arriving at the difference in order of magnitude. 
Figure~\ref{fig_bifurcacaoBaciaP3}(b) displays the attraction basins of the three listed periodic attractors for $\omega=4.25\pi$, 
corresponding to the point $P_3(\omega,\beta_1)=P_3(4.25\pi,0.2)$, 
shown in Fig.~\ref{fig_beta0Beta1Omega}(b). 
Initial conditions on plane $e_0\times i_0$ are in the same intervals and grid configuration as in Fig.~\ref{fig_baciaP2}. 
The color code used to identify the $3$ different basins is similar to the one employed in the bifurcation diagram, 
there is a slight difference in color tone to facilitate visualization. 
Pairs $(i_0,e_0)$ that lead to period $1$ orbit (with $i_{\rm max}\approx0.003$) 
are in black color; 
those leading to the period $3$ orbit (with $i_{\rm max}\approx0.012$) 
are in teal color; and 
those that evolve to period $2$ orbit (with $i_{\rm max}\approx0.030$) are in orange color. 
The gray region contains the pairs outside the normalized system domain (prohibited region). 
Notable are the alternating diagonal bands of the attraction basins. 
The highlight region in the bottom left corner, 
bounded by the white and magenta dashed box, 
is amplified in Fig.~\ref{fig_baciaP5zoom}(a) followed by two magnifications, 
in which the self-similarity and 
intricate nature of the attraction basins obtained for $P_3$ is evident. 

The succession of magnifications shown in Fig.~\ref{fig_baciaP5zoom} starts in the region marked in Fig.~\ref{fig_bifurcacaoBaciaP3}(b) and 
proceeds to scan intervals of initial conditions one order of magnitude smaller per panel, 
with the area covered in each successor being $10^{-2}$ times that of the previous one. 
%In all panels, 
%the plane is discretized in uniform grid of $1000\times1000$ points and 
Color code is the same as Fig.~\ref{fig_bifurcacaoBaciaP3}(b), 
identifying the $i_{\rm max}$ of the attractor resulting from each pair $(e_0,i_0)$.  
Table~\ref{tabela_baciaAtracao} presents the area percentage occupied by each basin of attraction for all $3$ evaluated ranges of initial conditions.  
\vspace{-0.1cm}
%
%--TABELA RECONFIGURADA
\renewcommand{\arraystretch}{1.5}
\newcommand{\PreserveBackslash}[1]{\let\temp=\\#1\let\\=\temp}
\newcolumntype{C}[1]{>{\PreserveBackslash\centering}m{#1}}
\begin{table}[!h]
	\centering
	\caption{Area percentage occupied by each basin of attraction for the $3$ evaluated regions. 
		First column identifies the panel in Fig.~\ref{fig_baciaP5zoom} and 
		the second column informs the respective range of initial conditions.} 
		%Last three columns sorted according to increasing $i_{\rm max}$.} 
	\begin{tabular}{ C{1.5cm} C{1.5cm} C{1.5cm} C{1.5cm} C{1.5cm}}
		\hline \hline
		\multicolumn{2}{c}{} & \multicolumn{3}{c}{$i_{\rm max}$} \\ \cline{3-5}
		\textbf{Panel} & $e_0,i_0\in$ & $0.003$   & $0.012$   & $0.030$   \\ \hline \hline
		a & $\left(0,10^{-1}\right]$  & $46.75\%$ & $47.78\%$ & $5.47\%$  \\ \hline
		b & $\left(0,10^{-2}\right]$  & $40,32\%$ & $44,02\%$ & $15,66\%$ \\ \hline
		c & $\left(0,10^{-3}\right]$  & $32,39\%$ & $65,97\%$ & $1,64\%$  \\ \hline \hline
	\end{tabular}
	\label{tabela_baciaAtracao}
\end{table}
\FloatBarrier

\FloatBarrier

First, 
in Fig~\ref{fig_baciaP5zoom}(a) we evaluate the initial conditions interval $0<e_0,i_0\leq10^{-1}$, 
where $46.75\%$ of the points lead to smallest $i_{\rm max}$ orbit and 
$5.47\%$ of them belong to the attraction basin of the periodic attractor with highest infectious peak. 
The box at the bottom left corner is enlarged in panel (b), 
where $0<e_0,i_0\leq0,10^{-2}$. 
In this region, 
the basin of attraction corresponding to $i_{\rm max}\approx0.003$ occupying $40,32\%$ of the total area and 
highest infectious peak attractor results from $15,66\%$ of the initial conditions. 
Panel (c) shows the last magnification, 
performed in the interval $0<e_0,i_0\leq0,10^{-3}$, 
being $65.97\%$ of the area occupied by the basin of period $3$ attractor, 
with the maximum of infectious $\approx0.012$. 
In this sample, 
only $1.64\%$ of the total area leads to $i_{\rm max}\approx0.030$. 
Given the discretization and 
intervals of both axes in Fig.~\ref{fig_baciaP5zoom}(c), 
one-point variation, 
being the increment of $10^{-6}$ either vertically or horizontally, 
represents a difference of $1$ individual in $1$ million of the host population for the initial condition of infectious or exposed. 
Resulting in significantly different maximum infectious values for this small change, 
with greater sensitivity for $e_0$ and 
$i_0$ of smaller orders of magnitude. 

\begin{figure}%[!b]
	\centering
	\includegraphics[width=0.92\columnwidth]{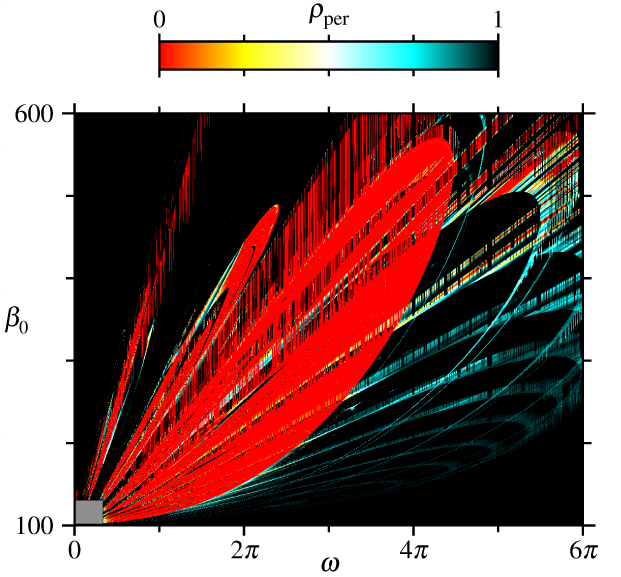}
	\caption{In colors the fraction $\rho_{\rm per}(\omega,\beta_0)$ of the trajectories that evolve to periodic dynamics, 
		from $100$ randomly assigned initial conditions. 
		Parameter plane discretized in uniform grid of $1000\times1000$ points, 
		with $0<\omega\leq6\pi$ and $100<\beta_0\leq600$. 
		Coexistence of periodic and 
		non-periodic orbits  ($0<\rho<1$) is observed in $13.20\%$ of the valid area. 
		In the gray square region, 
		the time series sampling was not enough to verify periodicity.} 
	\label{fig_probabilidadePeriodico}
\end{figure}

Finally, 
in Fig.~\ref{fig_probabilidadePeriodico} we obtain a fraction of initial conditions that result in periodic orbits for points along the parameter plane $\beta_0\times \omega$. 
We use the same intervals and 
parametric configuration of Fig.~\ref{fig_beta0Beta1Omega}(a). 
For each pair $(\omega,\beta_0)$, 
we draw $100$ equiprobable initial conditions in the interval $0<e_0,i_0<1$, 
respecting the constraint $e_0 + i_0\leq1$. 
Similar to what is made to obtain the bifurcation diagram in Fig.~\ref{fig_bifurcacaoBaciaP3}(a), 
starting from the $k$-th initial condition $C_k$, 
we evolve the system by $10^6$ integration steps even after the transient and 
evaluate only the last $7.5\times10^4$ trajectory points. 
In this section of the time series, 
we select the local maxima in the $s(t),~e(t)$ and $i(t)$ curves and 
check the periodicity with an accuracy of $10^{-6}$. 
Period $12$ is considered the maximum.
We do not focus on determining the period of each orbit, 
but just identify if it is periodic or not and, 
subsequently, 
compute the fraction $\rho_{\rm per}(\omega,\beta_0)$ of initial conditions that leads to periodic behavior. 
%for each pair $(\omega,\beta_0)$. 

%\FloatBarrier

In the sense explained above, 
Figure~\ref{fig_probabilidadePeriodico} consolidates information from $100$ parameter planes, 
each associated with the $k$-th initial condition for every pair $(\omega,\beta_0)$. 
We represent $\rho_{\rm per}$ by the color code. 
Pairs in this plane for which all evaluated initial conditions result in periodic orbits ($\rho_{\rm per}=1$) are in black color. 
The ones that $100\%$ led to non-periodic attractors ($\rho_{\rm per}=0$) are in red color. 
Intermediate cases are on the gradient from red to yellow ($\rho_{\rm per}=0.25$) and 
white ($\rho_{\rm per}=0.5$) and, 
on the other side, 
from white to cyan ($\rho_{\rm per}=0.75$) and black. 
In the small gray region (bottom left corner), 
the sample of time series are not enough to determine periodicity, 
or deny it, 
due to the high period of oscillations. 
%This way of identifying periodic behavior corroborates that of Sub-s.~\ref{sec_lyapunov}. 
Regions with some fraction of non-periodic behavior, 
where $\rho_{\rm per} < 1$, 
resemble the bands of $\lambda_1 > 0$ shown in Fig.~\ref{fig_beta0Beta1Omega}(a). 
Unlike the initial condition considered in Sub-s~\ref{sec_lyapunov}, 
here the exposed initial value is not null. 
Furthermore, 
accuracy, 
sampling and 
the maximum period adopted result in some differences between this and 
that parameter plane. 

The cyan bands in Fig.~\ref{fig_probabilidadePeriodico}, 
as well as the small yellow regions, 
demonstrate the coexistence of periodic and 
non-periodic orbits, 
related to seasonality parameters. 
Through this assessment, 
we obtain only periodic orbits for $61.50\%$ of the valid area (subtracting the gray region), 
exclusively non-periodic along $25.30\%$ and 
both behaviors coexist in $13.20\%$ of the parameter plane. 
%
%==================================================================================================
%
\section{CONCLUSION} \label{sec_conclusoes}
We numerically investigated the SEIRS model dynamics under a time-dependent transmission rate. 
Such temporal dependence is periodic and 
consists of expanding the concept of seasonality, 
considering different periods in addition to those that synchronize with the climatic seasons. 
Through parameter planes combining the seasonality frequency and 
typical epidemiological model parameters, 
we evidenced the chaotic dynamics occurrence, 
{referring to various analyses of reported data suggesting that some epidemics has chaotic behavior\cite{London1973,Olsen1990,Jones2021}.} 
Focusing on the transmission rate function,
we highlight the coexistence of chaotic and 
periodic orbits for certain parametric configurations, 
as well as a diversity of periodic attractors. 
We found that chaotic orbits may present lower infectious peaks than periodic ones, 
even so with the same temporal average of infectious cases. 
Assuming it is a disease that affects humans (which is not a premiss of the model), 
the predictability of periodic orbits facilitates the planning of public health campaigns.  
On the other hand, 
the lowest maximum number of cases in the chaotic attractor may represent a benefit when it comes to reduce the infection spread. 
{In terms of modeling real-world infections based on the system studied  in this work,  
	it is necessary consider the precision which the parameters can be determined, 
	since in the parameter planes there are narrow periodic bands immersed in chaotic regions.} 
{Also, 
	in the vicinity of the shrimp-like structures, 
	cascades of similar periodic regions occur, 
	entering in scales of very small parametric variations, 
	in such a way that small changes in the parameters can lead to a drastic change in the dynamic behavior.} 
%It is important to note that, for these results, the average transmission rate and the seasonality degree remained constant during system evolution.  

The system presents multistability of periodic orbits with different periods showing marked difference in the maximum infected values. 
By means of attraction basins, 
obtained for certain parametric configuration, 
we showed that small variations of the initial conditions can lead to different orders of magnitude of the maximum infectious agents number. 
Furthermore,
for a wide range of frequency and average transmissivity settings, 
orbits of periods up to $12$ coexist with larger periods and 
even non-periodic ones. 
Those two features of the system, 
present chaotic dynamics and 
multistability, 
lead to challenges for proposals of epidemic control, 
since chaotic dynamics reduce predictability and 
the multistable character can lead to significantly different periodic orbits through small changes of the initial conditions. 
The oscilation frequency of the transmission rate proved to be relevant to the system dynamics, 
being one of the determining factors for the occurrence of chaos. 
{Seasonality parameters also influence oscillations in the infectious curve in periodic scenarios, 
	resulting in different counts of local maxima within a period. 
	An investigation of this relationship can be carried out using isospike diagrams, 
	however, 
	is far from the focus of the present study, 
	so we consider it for a future work.}

%
%==================================================================================================
%
%\section*{AUTHOR'S CONTRIBUTIONS}

%
%==================================================================================================
%
\section*{ACKNOWLEDGMENTS}
The authors thank the financial support from the Brazilian Federal Agencies (CNPq); 
CAPES; 
Funda\-\c c\~ao A\-rauc\'aria. 
S\~ao Paulo Research Foundation (FAPESP) under Grant Nos. 2021/12232-0, 2018/03211-6, 2022/13761-9; 
R.L.V. received partial financial support from the following Brazilian government agencies: 
CNPq (403120/2021-7, 301019/2019-3), 
CAPES (88881.143103/2017-01), 
FAPESP (2022/04251-7). 
We thank 105 Group Science: \href{www.105groupscience.com}{www.105groupscience.com}. 
%
%==================================================================================================
%
\section*{DATA AVAILABILITY}
The data that support the findings of this study are available from the corresponding authors upon reasonable request.

%
%
%\newpage
\bibliography{references}
\end{document}